\documentclass[onecolumn,showpacs,preprintnumbers,amsmath,amssymb]{revtex4}
\usepackage{amssymb}
\usepackage[dvips]{graphicx}
\begin{document}

\title{Strong-coupling theory of high-temperature superconductivity and colossal magnetoresistance}
\author{A. S. Alexandrov}

\affiliation{Department of Physics, Loughborough University,
Loughborough, United Kingdom}

\large {
\begin{abstract} We argue that the extension of the BCS
theory to the strong-coupling regime describes the high-temperature
superconductivity of cuprates and the colossal magnetoresistance
(CMR) of ferromagnetic oxides if the phonon dressing of carriers and
strong attractive correlations are taken into account.   The
attraction between carriers, which is prerequisite to
high-temperature superconductivity, is caused by an almost
unretarted  electron-phonon interaction sufficient to overcome the
direct Coulomb repulsion in the strong-coupling limit, where
electrons become polarons and bipolarons (real-space electron or
hole pairs dressed by phonons).  The long-range Fr\"{o}hlich
electron-phonon interaction  has been identified as the most
essential in cuprates providing "superlight" lattice polarons and
bipolarons.  A number of key observations  have been predicted
and/or explained  with polarons and bipolarons including  unusual
isotope effects, normal state (pseudo)gaps, upper critical fields,
etc. Here  some kinetic, magnetic, and more recent thermomagnetic
normal state  measurements are interpreted in the framework of the
strong-coupling  theory, including the Nernst effect and normal
state diamagnetism. Remarkably, a similar strong-coupling approach
offers a simple explanation of CMR in ferromagnetic oxides, while
the conventional double-exchange (DEX) model, proposed half a
century ago and generalised more recently to include the
electron-phonon interaction, is in conflict with a number of modern
experiments. Among these experiments are site-selective
spectroscopies, which have shown that  oxygen p-holes are current
carriers rather than d-electrons in ferromagnetic manganites (and in
cuprates) ruling out DEX mechanism of CMR.  Also some samples of
ferromagnetic manganites manifest an insulating-like optical
conductivity at all temperatures contradicting the DEX notion that
their  ferromagnetic phase is metallic.
 On the other hand, the pairing of oxygen holes into heavy bipolarons
in the paramagnetic phase and their magnetic pair-breaking in the
ferromagnetic phase account for the first-order ferromagnetic phase
transition,  CMR,   isotope effects, and  pseudogaps in doped
manganites. Here we propose an explanation of  the phase coexistence
and describe the shape of  resistivity of manganites near the
transition   in the framework of the strong-coupling approach.
\end{abstract}


\maketitle

\section{INTRODUCTION: The ``Fr\"ohlich-Coulomb'' model}
\label{sect:intro}  

Although high-temperature superconductivity (HTS) has not yet been
targeted as `{\it the shame and despair of theoretical physics}', -
a label attributed to low-temperature superconductivity during the
first half-century after its discovery -  controversy of current
theoretical constructions has led many researchers to say that there
is no theory of HTS and no progress in understanding the phenomenon.
A significant fraction of theoretical research in the field   has
suggested that the interaction in novel superconductors is
essentially repulsive and unretarted, and it could provide high
$T_{c}$ without  phonons. Indeed strong  onsite repulsive
correlations (Hubbard $U$) are essential in shaping the insulating
state of  undoped (parent)  compounds.  Different from conventional
band-structure insulators with completely filled and empty Bloch
bands, the Mott insulator arises from a potentially metallic
half-filled band as a result of  the Coulomb blockade of electron
tunnelling to neighboring sites \cite{mott}. However,  the Hubbard
$U$ model shares an inherent difficulty in determining the order
when the Mott-Hubbard insulator is doped. While some groups have
claimed that it describes high-$T_{c}$ superconductivity at finite
doping, other authors could not find any superconducting
instability. Therefore it has been concluded that models of this
kind are highly conflicting and confuse the issue by exaggerating
the magnetism rather than clarifying it \cite{lau}.

 The Hubbard-$U$ model of high
temperature superconductivity or its strong-coupling $"t-J"$
approximation   are  also refutable  on the experimental ground. The
characteristic magnetic interaction, which is allegedly responsible
for the pairing in the  model, is the spin exchange interaction,
$J=4t^{2}/U$, of the order of $0.1$ eV (here $t$ is the hopping
integral). On the other hand, a simple parameter-free estimate of
the Fr\"ohlich electron-phonon interaction (routinely neglected
within the Hubbard $U$ approach) yields the effective attraction as
high as $1$ eV \cite{alebook}. This estimate  is obtained using the
familiar expression for the polaron level shift, $E_{p},$  the
high-frequency, $\epsilon _{\infty }$, and the static, $\epsilon
_{0},$ dielectric constants of the host insulator, measured
experimentally \cite{alebra},
\begin{equation}
E_{p}={\frac{1}{{2\kappa }}}\int_{BZ}{\frac{d^{3}q}{{(2\pi )^{3}}}}{\frac{%
4\pi e^{2}}{{q^{2}}}},
\end{equation}
where $\kappa ^{-1}=\epsilon _{\infty }^{-1}-\epsilon _{0}^{-1}$ and
the size of the integration region is the Brillouin zone (BZ). Since
$\epsilon _{\infty }=5$, $\epsilon_0=30$ in La$_2$CuO$_4$ and
$\epsilon _{\infty }=3.9$, $\epsilon_0=16$ in LaMnO$_3$ one obtains
$E_p=0.65$ eV and $E_p=0.88$ eV in La$_2$CuO$_4$ and LaMnO$_3$,
respectively. Hence the attraction, which is about $2E_p$, induced
by the lattice deformation in cuprates and manganites is  one order
of magnitude larger than the exchange (magnetic) interaction.  There
is virtually no screening of
 e-ph interactions with $c-$axis polarized optical phonons in
cuprates because the upper limit for the out-of-plane plasmon
frequency ($< 200$ cm$^{-1}$)\cite{mar} is well below the
characteristic phonon frequency, $\omega\approx$ 400 - 1000 cm
$^{-1}$ .  The screening in manganites is also very poor since the
mobility of carriers is very low. As a result of poor screening the
magnetic interaction remains small compared with the Fr\"ohlich
interaction at any doping. Further compelling evidence for the
strong e-ph interactions has come from the isotope effects in
cuprates \cite{ZHAO} and manganites \cite{zhao1}, recent high
resolution angle resolved photoemission spectroscopies
 \cite{LAN}, and a number of earlier optical \cite{mic,ita} and neutron-scattering \cite{ega} studies.
  Hence any realistic approach to HTS
 in cuprates, other doped oxides and fullerenes, and to CMR in ferromagnetic oxides should treat the long-range Coulomb and
\emph{unscreened} e-ph interactions on an equal footing.

In the past decade we have developed a "Fr\"ohlich-Coulomb" model
(FCM) \cite{ALEXAND,alekor,alebook} to deal with the strong
long-range Coulomb and
 e-ph interactions in cuprates, manganites and other related compounds.
The model Hamiltonian explicitly includes a long-range
electron-phonon and the Coulomb interactions as well as the kinetic
and deformation energies.  The implicitly present large Hubbard $U$
term prohibits double occupancy and removes the need to distinguish
fermionic spins since the exchange interaction is negligible
compared with the direct Coulomb and the electron-phonon
interactions. The model also provides a simple explanation of CMR in
ferromagnetic oxides if the exchange interaction of p-holes with
d-electron spins is included in the Hamiltonian \cite{alebra2} (see
below). Introducing spinless fermionic, $c_{\bf n}$, and phononic,
$d_{{\bf m}\alpha }$, operators the Hamiltonian of the model  is
written as
\begin{eqnarray}
H = & - & \sum_{\bf n \neq n'} \left[ t({\bf n-n'}) c_{\bf
n}^{\dagger } c_{\bf n'} - V_{c}({\bf n-n'}) c_{\bf n}^{\dagger}
c_{\bf n}c_{\bf n'}^{\dagger } c_{\bf n'} \right]  \nonumber \\
& - &  \sum_{\bf n m} \omega_{\alpha} g_{\alpha}({\bf m-n}) ({\bf
e}_{\alpha } \cdot {\bf u}_{\bf m-n}) c_{\bf n}^{\dagger } c_{\bf
n} (d_{{\bf m}\alpha}^{\dagger}+d_{{\bf m}\alpha }) \nonumber \\
&+&
 \sum_{{\bf m}\alpha} \omega_{\alpha}\left( d_{{\bf m}\alpha
}^{\dagger} d_{{\bf m}\alpha }+1/2 \right),
\end{eqnarray}
where ${\bf e}_{ \alpha}$ is the polarization vector of the
$\alpha$th vibration coordinate, ${\bf u}_{\bf m-n} \equiv ({\bf
m-n})/|{\bf m-n}|$ is the unit vector in the direction from electron
${\bf n}$ to  ion ${\bf m}$, $g_{\alpha}({\bf m-n)}$ is the
dimensionless e-ph coupling function, and $V_{c}({\bf n-n'})$ is the
inter-site Coulomb repulsion. $g_{\alpha}({\bf m-n)}$ is
proportional to the {\em force} acting between the electron on site
${\bf n}$ and the ion on ${\bf m}$. For simplicity, we assume that
all the phonon modes are non-dispersive with the frequency
$\omega_{\alpha}$. We also use $\hbar =k_B=c=1$.

The  Hamiltonian, Eq.(2), has been solved analytically  by using the
$"1/\lambda"$ multi-polaron expansion technique \cite{alebook} in
the strong limit where the e-ph coupling constant is large, $\lambda
=E_p/zt >1$. Here the polaron level shift is $ E_{p} = \sum_{{\bf n}
\alpha} \omega_{\alpha} g_{\alpha}^{2}({\bf n}) ({\bf
e}_{\alpha}\cdot {\bf u}_{\bf n})^{2} ,$  and $zt$ is a
half-bandwidth in the rigid lattice. The model shows a reach phase
diagram depending on the ratio of the inter-site Coulomb repulsion
$V_{c}$ and the polaron level shift $E_{p}$ \cite{alekor}. The
ground state of FCM is a \emph{polaronic} Fermi liquid when the
Coulomb repulsion is large, a \emph{bipolaronic} high-temperature
superconductor at  intermediate Coulomb repulsions, and a
charge-segregated insulator if the repulsion is weak. FCM predicts
\emph{superlight } polarons and bipolarons in cuprates with a
remarkably  high superconducting critical temperature. Cuprate
bipolarons are relatively light because they are $inter-site$ rather
than  $on-site$ pairs  due to the strong on-site repulsion,  and
because mainly $c$-axis polarized optical phonons are responsible
for the in-plane mass renormalization. The relatively small mass
renormalization of polaronic and bipolaronic carries in FCM has been
confirmed numerically using the exact QMC \cite{Korn2}, cluster
diagonalization \cite{feh3} and variational \cite{bon2} simulations.

(Bi)polarons describe many properties of
 cuprates \cite{alebook}, in particular  normal-state transport properties (section 2),
 the  Nernst effect (section 3), and the normal state diamagnetism (section 4).
The strong-coupling theory also provides  an explanation for the
phase separation and coexistence  and describes the shape of
resistive and magnetic transitions in manganites (section 5).

\section{Normal state in-plane resistivity, Hall effect and magnetic susceptibility of cuprates in the bipolaron model}

The low-energy FCM electronic structure  of  cuprates  is shown in
Fig.1 \cite{MOTT}. Polaronic p-holes  are bound into lattice
inter-site singlets (A) or into singlets and  triplets (B) (if spins
are included in Eq.(2)) at any temperature. Above T$_{c}$ a charged
bipolaronic  Bose-liquid is non-degenerate and below $T_{c}$ phase
coherence (ODLRO) of the preformed bosons sets in. The state above
$T_{c}$ is perfectly "normal" in the sense that the off-diagonal
order parameter (i.e. the Bogoliubov-Gor'kov anomalous average
$\cal{F}(\mathbf{r,r^{\prime }})=\langle \psi_{\downarrow
}(\mathbf{{r})\psi _{\uparrow }({r^{\prime }}\rangle}$) is zero
above the resistive transition temperature $T_{c}$. Here
$\psi_{\downarrow,\uparrow }(\mathbf{r})$ annihilates  electrons
with spin $\downarrow, \uparrow$ at point ${\bf r}$. Triplet and
singlet states are separated by the exchange energy $J$ which
explains the spin gap observed in a number of NMR and neutron
scattering experiments. There are also thermally excited single
polarons in the model. Their density becomes comparable with the
bipolaron density at the temperature $T^{\ast}$ which is about half
of the bipolaron binding energy $\Delta $, in accordance with the
experimentally observed crossover regime at $T^{\ast} > T_c$ and the
normal state pseudogaps in  cuprates.
\begin{figure}
\begin{center}
\includegraphics[angle=-0,width=0.45\textwidth]{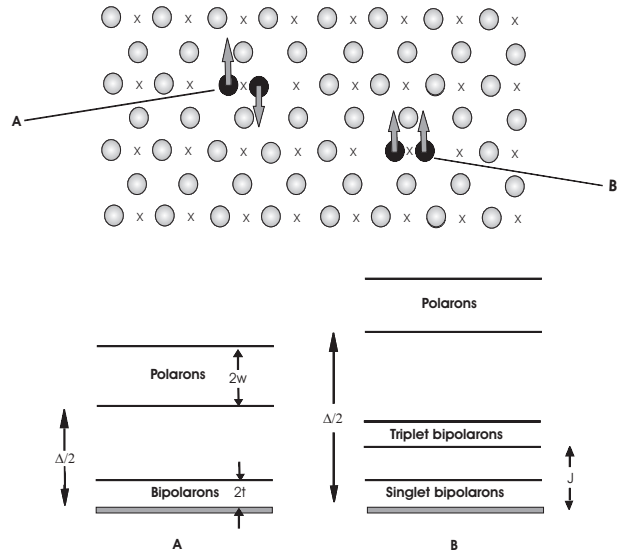}
\vskip -0.5mm \caption{Bipolaron picture of high temperature
superconductors. $A$ corresponds to the singlet intersite bipolaron.
$B$ is the triplet intersite bipolaron, which naturally includes the
addition of an extra excitation band. The crosses are copper sites
and the circles are oxygen sites, $w$ is a half bandwidth of the
polaron band, $t$ is a half bandwidth of the bipolaronic band,
$\Delta/2$ is the bipolaron binding energy per polaron and $J$ is
the exchange energy per bipolaron.}
\end{center}
\end{figure}

A nonlinear temperature dependence of the $in$-plane resistivity
below $T^{\ast }$, a temperature-dependent paramagnetic
susceptibility, and a peculiar maximum in the Hall ratio well above
$T_c$ have remained  long-standing problems of cuprate physics. The
bipolaron model provides their quantitative description
\cite{alezavdzu}. Thermally excited phonons and (bi)polarons are
well decoupled in the strong-coupling regime of the electron-phonon
interaction \cite{alebook}, so
 the conventional Boltzmann kinetics for mobile polaronic and bipolaronic carries is
applied. Here we use a `minimum' bipolaron model Fig.1A, which
includes the singlet bipolaron band and the spin 1/2 polaron band
separated by $T^{\ast }$, and the $\tau -$approximation   in weak
electric ${\bf E}$ and  magnetic fields, $\bf B\perp E$.
\begin{figure}
\begin{center}
\includegraphics[angle=-0,width=0.30\textwidth]{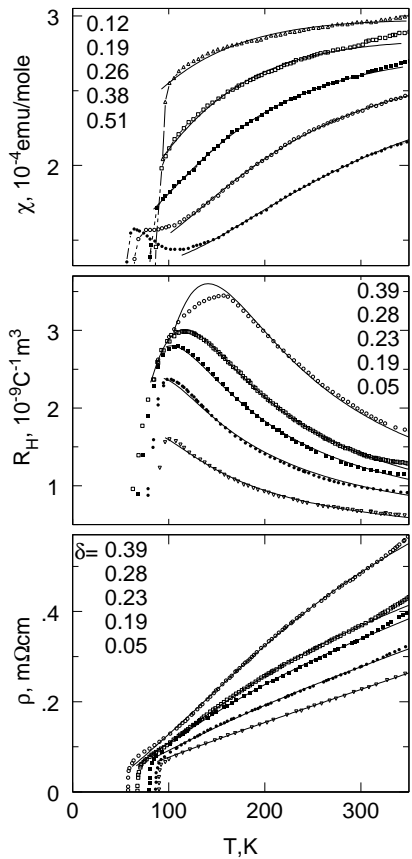}
\vskip -0.5mm
\caption{Uniform magnetic susceptibility, $\chi(T)$, Hall ratio, $R_H(T)$ and resistivity, $\rho(T)$, of underdoped $YBa_{2}Cu_{3}O_{7-%
\delta }$ fitted by the theory; see the Table for parameters. }
\end{center}
\end{figure}

Bipolaron and single-polaron non-equilibrium distributions are found
as
\begin{equation}
f({\bf k})=f_{0}(E)+\tau \frac{\partial f_{0}}{\partial E}{\bf v}\cdot
\left\{ {\bf F}+\Theta {\bf n}\times {\bf F}\right\} ,  \label{4}
\end{equation}
where ${\bf v=}\partial E/\partial {\bf k,}$ ${\bf
F}={\vec{\bf\nabla}}(\mu -2e\phi )$, $f_{0}(E)=[y^{-1}{\exp
(E/T)-1]}^{-1}$ and the Hall angle $\Theta =\Theta _{b}=2eB\tau
_{b}/m_{b}$ for bipolarons with the energy $E=k^{2}/(2m_{b})$, and
${\bf F}={\vec{\bf\nabla}}(\mu /2-e\phi ) $,
$f_{0}(E)=\{y^{-1/2}{\exp [(E+T^{\ast })/T]+1\}}^{-1}$,
$E=k^{2}/(2m_{p})$, and $\Theta =\Theta _{p}=eB\tau
_{p}/m_{p}$ for thermally excited polarons. Here
$m_{b}$ and $m_p$ are the bipolaron and polaron mass, respectively, $y=\exp
(\mu /T),$ $\mu $ is the chemical potential,  and ${\bf %
n=B/}B$ is a unit vector in the direction of the magnetic field.
Eq.(3) is used to calculate the electrical resistivity and the Hall
ratio as
\begin{eqnarray}
\rho &=&\frac{m_{b}}{4e^{2}\tau _{b}n_{b}(1+An_{p}/n_{b})}, \\
R_{H} &=&\frac{1+2A^{2}n_{p}/n_{b}}{2en_{b}(1+An_{p}/n_{b})^{2}},
\end{eqnarray}
where $A=\tau _{p}m_{b}/(4\tau _{b}m_{p})$. The atomic densities of
quasi two-dimensional carriers are found as
\begin{eqnarray}
n_{b}=\frac{m_{b}T}{2\pi }|\ln (1-y)|, \\
n_{p}=\frac{m_{p}T}{\pi }\ln \left[ 1+y^{1/2}\exp \left( -T^{\ast }/T\right) %
\right] .
\end{eqnarray}
and the chemical potential is determined by doping $x$ using
$2n_{b}+n_{p}=x-n_{L}$,
where $n_{L}$ is the number of carriers localised by disorder (here we take
the lattice constant $a=1$).

Polarons are not degenerate. Their number  remains small compared
with twice the number of bipolarons, $n_p/(2n_b)<0.2$, in the
relevant temperature range $T < T^{\ast }$, so that
\begin{equation}
 y\approx 1-\exp(-T_0/T),
\end{equation}
where $T_0=\pi (x-n_L)/m_b \approx T_c$ is about the superconducting
critical temperature of the (quasi)two-dimensional Bose gas. Because
of this reason, the experimental $T_c$ was taken as  $T_0$ in our
fits. Using Eqs.(7,6,5) we obtain
\begin{equation}
R_{H}(T)=R_{H0}\frac{1+2A^2y^{1/2}(T/T_{c})\exp \left( -T^{\ast
      }/T\right) }{
[1+A(T/T_{c})y^{1/2}\exp \left( -T^{\ast }/T\right) ]^{2}},
\end{equation}
where $R_{H0}=[e(x-n_{L})]^{-1}$.  If we assume that the number of
localised carriers depends only weakly on temperature in underdoped
cuprates since their average  ionisation energy is sufficiently
large, then $R_{H0}$ is temperature independent at $T<T^{\ast }$. As
proposed in Ref.\cite{BRAT} the scattering rate  at relatively high
temperatures is due to inelastic
 collisions of itinerant carriers with those localised by
 disorder, so it is proportional to $T^2$. We also have to take into account the residual scattering of polarons off optical
 phonons, so that
$\tau^{-1}=aT^{2}+b\exp{(-\omega /T)}$, if the temperature is low
compared with the characteristic phonon energy $\omega$. The
relaxation times of each type of carriers scales with their charge
$e^\ast$ and mass as $\tau_{p,b} \propto
m_{p,b}^{-3/2}(e^\ast)^{-2}$, so we estimate $A=(m_b/m_p)^{5/2}
\approx 6$ if we take $m_{b}\approx 2m_p$ . As a result  the
in-plane resistivity is given by
\begin{equation}
\rho (T)=\rho _{0}\frac{(T/T_{1})^{2}+\exp \left( -\omega /T\right) }{[1+A(T/T_{c})y^{1/2}\exp \left( -T^{\ast }/T\right) ]},
\end{equation}
where $\rho _{0}=bm_{b}/[2e^{2}(x-n_L)]$ and $T_{1}=(b/a)^{1/2}$ are
temperature independent. Finally, one can easily obtain the uniform magnetic
susceptibility due to nondegenerate spin 1/2 polarons as \cite{AKM}
\begin{equation}
\chi (T)=By^{1/2}\exp \left( -T^{\ast }/T\right) +\chi _{0},
\end{equation}
where $B=(\mu _{B}^{2}m_{p}/\pi )$, and $\chi_{0} $ is the magnetic susceptibility of the parent Mott insulator.
\vspace{5mm}

\begin{tabular}{|c|c||c|c|c||c|c|c|c|c|}

 \hline
$\delta$ &$T_c$& $\rho_0$&$R_{H0}$ &$10^4B$ & $10^4\chi_0$ & $T^*$& $\omega$  & $T_1$   \\
& K & $m\Omega cm$&$\frac{10^{-9}m^3}{C}$& $\frac{emu}{mole}$ &
$\frac{emu}{mole}$ &K&K  &K

\\ \hline
  0.05 &  90.7 & 1.8  & 0.45&&& 144&447 & 332
\\ \hline
  0.12 &   93.7 &&&  2.6 &2.1&155 &&
\\ \hline
  0.19 &  87 &3.4 & 0.63 &4.5&1.6& 180&477& 454
\\ \hline
  0.23 &  80.6& 5.7 & 0.74&&&210& 525  & 586
\\ \hline
  0.26 &   78 & & &5.4 &1.5&259&&
\\ \hline
  0.28 & 68.6 & 8.9& 0.81&&&259& 594  & 786
\\ \hline
  0.38 & 61.9 && &7.2 &1.4& 348&&
\\ \hline
  0.39 & 58.1& 17.8& 0.96&&&344& 747 & 1088
\\ \hline
  0.51 & 55 && & 9.1 &1.3& 494&&
\\ \hline
\end{tabular}

The present model numerically fits the Hall ratio, $R_H(T)$, the in-plane resistivity, $\rho(T)$, and the
magnetic susceptibility $\chi(T)$ of $YBa_{2}Cu_{3}O_{7-\delta }$ within the physically relevant range of all parameters (see Fig. 2 and the Table). The ratio of polaron and
bipolaron mobilities $A=7$ used in all fits is close to the above
estimate,  and $\chi _{0}\approx 1.5\times 10^{-4}emu/mole$ is very close
to the susceptibility of a slightly doped insulator
\cite{coop}. The
maximum of $R_{H}(T)$ is due to the contribution of thermally excited polarons
into transport, and the temperature dependence of the in-plane resistivity
below $T^{\ast }$ is due to this contribution and the combination of the
carrier-carrier and carrier-phonon scattering. The
characteristic phonon frequency from the resistivity fit (Table) decreases
with doping
and the pseudogap $T^{\ast }$ shows the doping behaviour as observed in
other independent experiments.

Notwithstanding our explanation of the Hall ratio, the in-plane
resistivity and the bulk magnetic susceptibility might be not so
convincing as a direct measurement of the double
 charge $2e$ on carriers in the normal state. In
1993, we  discussed the thermal conductivity of preformed bosons
\cite{NEV}. The contribution from  carriers to the thermal transport
provided by the Wiedemann-Franz law depends strongly on the
elementary charge as $\sim(e^{\ast })^{-2}$ and should be
significantly suppressed if $e^{\ast}=2e$. The Lorenz number, $L$,
has been directly measured in $YBa_{2}Cu_{3}O_{6.95}$ by Zhang et
al. \cite{zha} using the thermal Hall conductivity. Remarkably, the
measured value of $L$ just above $T_{c}$ was found just the same as
predicted by the bipolaron model \cite{NEV}, $L\approx 0.15L_{e}$,
where $L_{e}$ is the conventional Fermi-liquid Lorenz number. The
breakdown of the Wiedemann-Franz law has been also explained in the
framework of the bipolaron model \cite{leeale}.

\section{Normal-state Nernst effect}

In disagreement with the weak-coupling BCS  and the strong-coupling
 bipolaron theories a significant fraction of research in the field
of high-temperature superconductivity suggests that the
superconducting transition is only a phase ordering while the
superconducting order parameter $\cal{F}(\mathbf{r,r^{\prime }})$
remains nonzero above the resistive $T_c$. One of the key
experiments supporting this viewpoint is the large Nernst signal
observed  in the normal (i.e. resistive) state of cuprates  (see
Ref. \cite{xu,cap,cap2} and references therein). Some authors
\cite{xu,ong}  claim that numerous resistive determinations of the
upper critical field, $H_{c2}(T)$ in cuprates have been misleading
since the Nernst signal \cite{xu} and the diamagnetic magnetization
\cite{ong} imply that  $H_{c2}(T)$  remains large at $T_c$ and
above. They propose a "vortex scenario", where the long-range phase
coherence is destroyed by mobile vortices, but the amplitude of the
off-diagonal order parameter remains finite and  the Cooper pairing
with a large binding energy exists  well above $T_c$ supporting the
so-called  "preformed Cooper-pair" or "phase fluctuation"  model
\cite{kiv}. The model is  based on the assumption that the
superfluid density is small compared with  the normal carrier
density in cuprates.  These interpretations  seriously undermine
many theoretical and experimental  works on superconducting
cuprates, which consider the state above $T_c$ as perfectly normal
with no off-diagonal order, either long or short.

We believe that  the vortex (or phase fluctuation) scenario
contradicts  straightforward resistive and other measurements, and
it is theoretically inconsistent. This scenario is impossible to
reconcile with the extremely sharp resistive transitions at $T_c$ in
high-quality underdoped, optimally doped  and overdoped cuprates.
For example, the in-plane and out-of-plane resistivity of $Bi-2212$,
where the anomalous Nernst signal has been measured \cite{xu}, is
perfectly "normal" above $T_c$, Fig.3, showing only a few percent
positive or negative magnetoresistance \cite{zavale}.
\begin{figure}
\begin{center}
\includegraphics[angle=-0,width=0.40\textwidth]{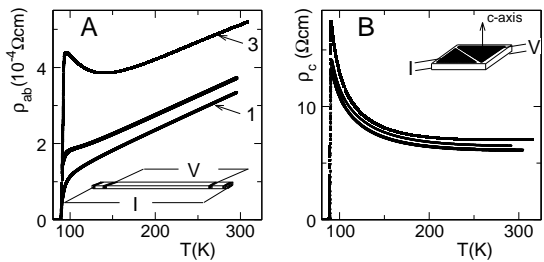}
\vskip -0.5mm \caption{In-plane (A) and out-of-plane (B) resistivity
of 3   single crystals of Bi$_2$Sr$_2$CaCu$_2$O$_8$ \cite{zavale}
showing no signature of phase fluctuations well above the resistive
transition temperature. }
\end{center}
\end{figure}
Both in-plane \cite{buc,mac0,boz,fra,gan} and out-of-plane \cite
{alezavnev,out,out2} resistive transitions  of high-quality samples
are sharp and remain sharp in the magnetic field providing a
reliable determination of the genuine $H_{c2}(T)$. The vortex
entropy \cite{cap} estimated from the Nernst signal is an order of
magnitude smaller than the difference between the entropy of the
superconducting state and the extrapolated entropy of the normal
state obtained from the specific heat. The preformed Cooper-pair
model \cite{kiv}  is incompatible with a great number of
thermodynamic, magnetic, and kinetic measurements, which show that
only holes (density x), doped into a parent insulator are carriers
\emph{both} in  the normal and the superconducting states of
cuprates. The assumption \cite{kiv} that the superfluid density is
small compared with the normal-state carrier density is also
inconsistent with the theorem \cite{leg}, which proves that the
number of supercarriers at $T=0$K  should be the same as the number
of normal-state carriers in any  clean superfluid.

Recently we described the unusual Nernst signal in cuprates in a
different manner as the  normal state phenomenon \cite{alezav}. Here
we extend our description to cuprates with very low doping level
accounting  for their Nernst signal, the thermopower and the
insulating-like in-plane low temperature resistance
\cite{xu,cap,cap2}.

Thermomagnetic effects appear in conductors subjected to a
longitudinal temperature gradient $\nabla _{x}T$ in $x$ direction
and a perpendicular magnetic field  in $z$ direction. The transverse
Nernst-Ettingshausen effect \cite{nernst}  (here the Nernst effect)
is the appearance of a transverse electric field $E_y$ in the third
direction. When bipolarons are formed in the strong-coupling regime,
the chemical potential is  negative, Eq.(8). It is found in the
impurity band just below the mobility edge at $T>T_c$. Carriers,
localised  below the mobility edge contribute to the longitudinal
transport together with the itinerant carriers in extended states
above the mobility edge. Importantly the contribution of localised
carriers of any statistics to the \emph{ transverse} transport is
normally small \cite{ell} since a microscopic Hall voltage will only
develop at junctions in the intersections of the percolation paths,
and it is expected that these are few for the case of hopping
conduction among disorder-localised states \cite{mott2}. Even if
this contribution is not negligible, it adds  to the contribution of
itinerant carriers to produce a large Nernst signal,
$e_{y}(T,B)\equiv -E_{y}/\nabla _{x}T$, while it reduces the
thermopower $S$ and the Hall angle $\Theta$. This unusual "symmetry
breaking" is completely at variance with  ordinary metals where the
familiar "Sondheimer" cancelation \cite{sond} makes  $e_{y}$ much
smaller than $S\tan \Theta$ because of the electron-hole symmetry
near the Fermi level. Such  behaviour originates in the "sign" (or
"$p-n$") anomaly of the Hall conductivity of localised carriers. The
sign of their Hall effect  is often $opposite$ to that of the
thermopower as observed in many amorphous semiconductors \cite{ell}
and described theoretically \cite{fri}.

The Nernst signal is expressed in terms of the kinetic coefficients
$\sigma _{ij}$ and $\alpha _{ij}$ as
\begin{equation}
e_{y}={\frac{{\sigma
_{xx}\alpha _{yx}-\sigma _{yx}\alpha _{xx}}}{{\sigma
_{xx}^{2}+\sigma _{xy}^{2}}}},
\end{equation}
where the current density  is given by $j_{i}=\sigma
_{ij}E_{j}+\alpha _{ij}\nabla _{j}T$.
 When the chemical potential $\mu$ is at the mobility edge,  the localised carriers contribute to the transport,
 so  $\sigma _{ij}$ and $\alpha _{ij}$ in Eq.(12) can be expressed
as $\sigma^{ext} _{ij}+\sigma^{l}_{ij}$ and $\alpha^{ext}
_{ij}+\alpha^{l}{ij}$, respectively. Since the Hall mobility of
carriers localised below $\mu$, $\sigma^{l}_{yx}$, has the  sign
opposite to that of carries in the extended states above $\mu$,
$\sigma^{ext}_{yx}$, the sign of the off-diagonal Peltier
conductivity $\alpha^{l}_{yx}$ should be the same as the sign of
$\alpha^{ext}_{yx}$. Then  neglecting the magneto-orbital effects in
the resistivity (since $\Theta \ll 1$ \cite{xu}) we obtain
\begin{equation}
S\tan \Theta \equiv {\sigma _{yx}\alpha _{xx}\over{\sigma
_{xx}^{2}+\sigma _{xy}^{2}}} \approx\rho (\alpha ^{ext}_{xx}-|\alpha ^{l}_{xx}|) (\Theta^{ext}-|\Theta^{l}|)
\end{equation}
and
\begin{equation}
e_{y}\approx\rho (\alpha^{ext} _{yx}+|\alpha^{l} _{yx}|)-S\tan \Theta,
\end{equation}
where $\Theta^{ext}\equiv \sigma^{ext}_{yx}/\sigma_{xx}$, $\Theta^{l}\equiv \sigma^{l}_{yx}/\sigma_{xx}$, and $\rho=1/\sigma_{xx}$ is the resistivity.

\begin{figure}
\begin{center}
\includegraphics[angle=-90,width=0.40\textwidth]{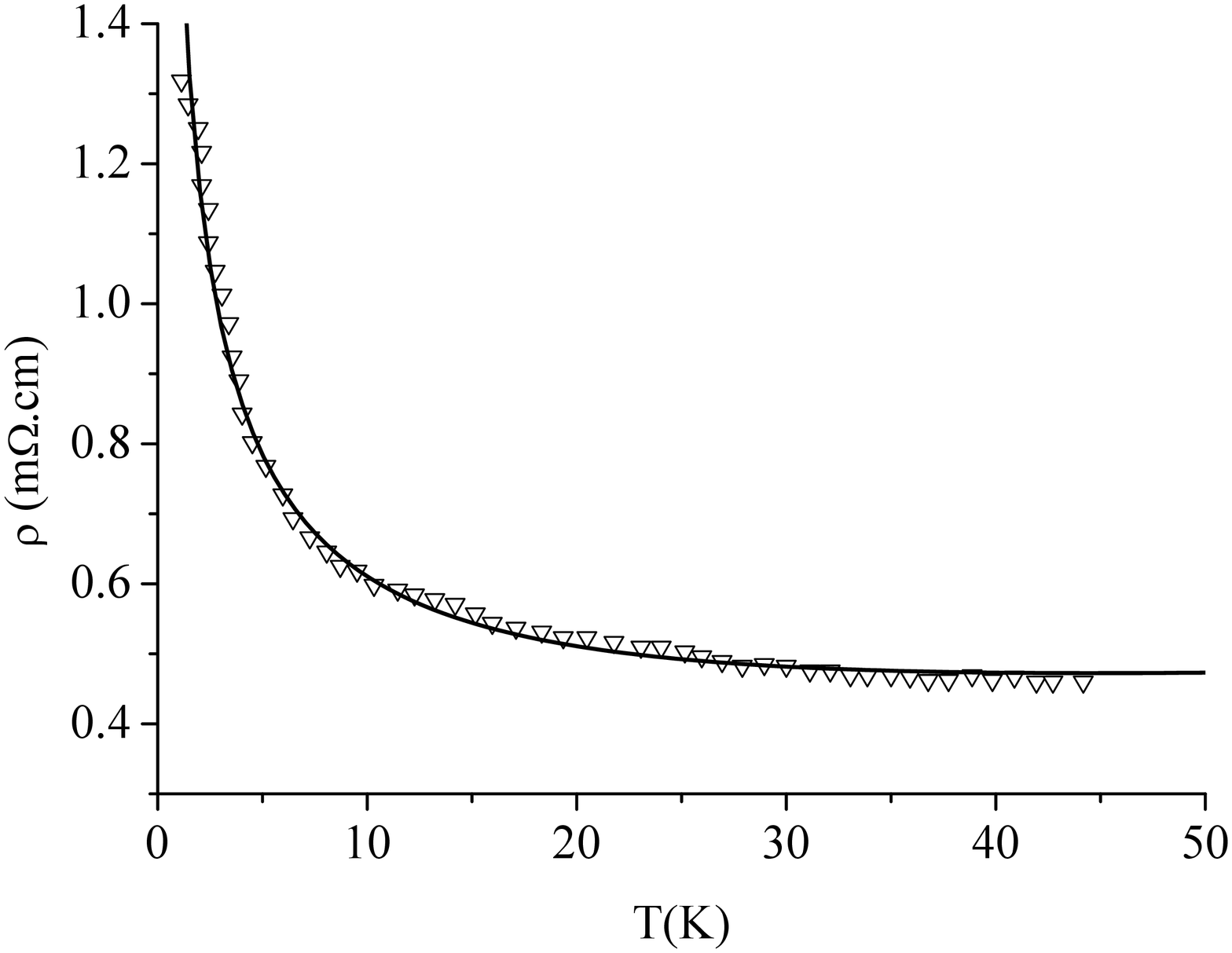}
\vskip -0.5mm \caption{Normal state in-plane resistivity of
underdoped La$_{1.94}$ Sr$_{0.06}$CuO$_4$ (triangles \cite{cap}) as
revealed in the field $B=12$ Tesla  and compared with the bipolaron
theory, Eq.(17) (solid line).}
\end{center}
\end{figure}
Clearly the model, Eqs.(13,14) can account for a low value of
$S\tan\Theta$ compared with a large value of $e_y$ in some
underdoped cuprates \cite{xu,cap2} due to the sign anomaly. Even in
the case when  localised bosons contribute little to the
conductivity   their contribution to the thermopower $S=\rho (\alpha
^{ext}_{xx}-|\alpha ^{l}_{xx}|))$ could almost cancel  the opposite
sign contribution of itinerant carriers \cite{alezav}. Indeed the
longitudinal conductivity of itinerant two-dimensional bosons,
$\sigma^{ext} \propto \int_0 dE E df(E)/dE$ diverges logarithmically
when $\mu$ in the Bose-Einstein distribution function
$f(E)=[\exp((E-\mu)/T)-1]^{-1}$ goes to zero and the relaxation time
$\tau$ is a constant. At the same time $\alpha^{ext}_{xx}\propto
\int_0 dE E(E-\mu) df(E)/dE$ remains finite, and it could have the
magnitude comparable   with  $\alpha^{l}_{xx}$. Statistics of
bipolarons  gradually changes from Bose to Fermi statistics with
lowering energy across the mobility edge because of the Coulomb
repulsion of bosons in localised states  \cite{alegile}. Hence one
can use the same expansion near the mobility edge as in  ordinary
amorphous semiconductors to obtain the familiar textbook result
$S=S_0T$ with a constant $S_0$ at low temperatures \cite{mott3}. The
model becomes particularly simple, if we   neglect the localised
carrier contribution to $\rho$, $\Theta$ and $\alpha_{xy}$, and take
into account that $\alpha^{ext}_{xy} \propto B/\rho^2$ and
$\Theta^{ext}\propto B/\rho$ in accordance with the Boltzmann
theory. Then Eqs.(13,14) yield
\begin{equation}
S\tan \Theta  \propto T/\rho
\end{equation}
and
\begin{equation}
e_{y}(T,B)\propto (1-T/T_1)/\rho.
\end{equation}
According to our earlier suggestion \cite{alelog} the
insulating-like low-temperature dependence of $\rho(T)$ in
underdoped cuprates  originates from the elastic scattering of
nondegenerate itinerant carriers off charged  impurities. As in
section 2 we assume here that the carrier density is temperature
independent at low temperatures in agreement  with the
temperature-independent Hall effect \cite{per}. The relaxation time
of nondegenerate carriers  depends on temperature as $\tau \propto
T^{-1/2}$ for scattering off short-range deep potential wells, and
as $T^{1/2}$ for very shallow wells \cite{alelog}. Combining both
scattering rates we obtain
\begin{equation}
\rho =\rho_0[(T/T_2)^{1/2}+(T_2/T)^{1/2}].
\end{equation}
Eq.(17) with $\rho_0=0.236$ m$\Omega\cdot$cm and $T_2=44.6$K fits
extremely well the experimental insulating-like normal state
resistivity of underdoped La$_{1.94}$ Sr$_{0.06}$CuO$_4$ in the
whole low-temperature range from  2K up to 50K, Fig.4,  as revealed
in the field $B=12$ Tesla \cite{cap,cap2}. Another high quality fit
can be  obtained combining the Brooks-Herring formula for the 3D
scattering off charged impurities, as proposed in Ref.\cite{kast}
for almost undoped $LSCO$, or the Coulomb scattering in 2D ($\tau
\propto T$) and a temperature independent scattering rate off
neutral impurities with the carrier exchange \cite{erg} similar to
the scattering of slow electrons by hydrogen atoms. Importantly our
expressions (15,16) for $S\tan \Theta$ and $e_y$ do not depend on
the particular scattering mechanism. Taking into account the
excellent fit of Eq.(17) to the experiment, they can be
parameterized as
\begin{equation} S\tan \Theta = e_0
{(T/T_2)^{3/2}\over{1+T/T_2}},
\end{equation}
and
\begin{equation}
e_{y}(T,B)=e_0{(T_1-T) (T/T_2)^{1/2}\over{T_2+T}} ,
\end{equation}
where $T_1$ and $e_0$ are temperature independent.
\begin{figure}
\begin{center}
\includegraphics[angle=270,width=0.50\textwidth]{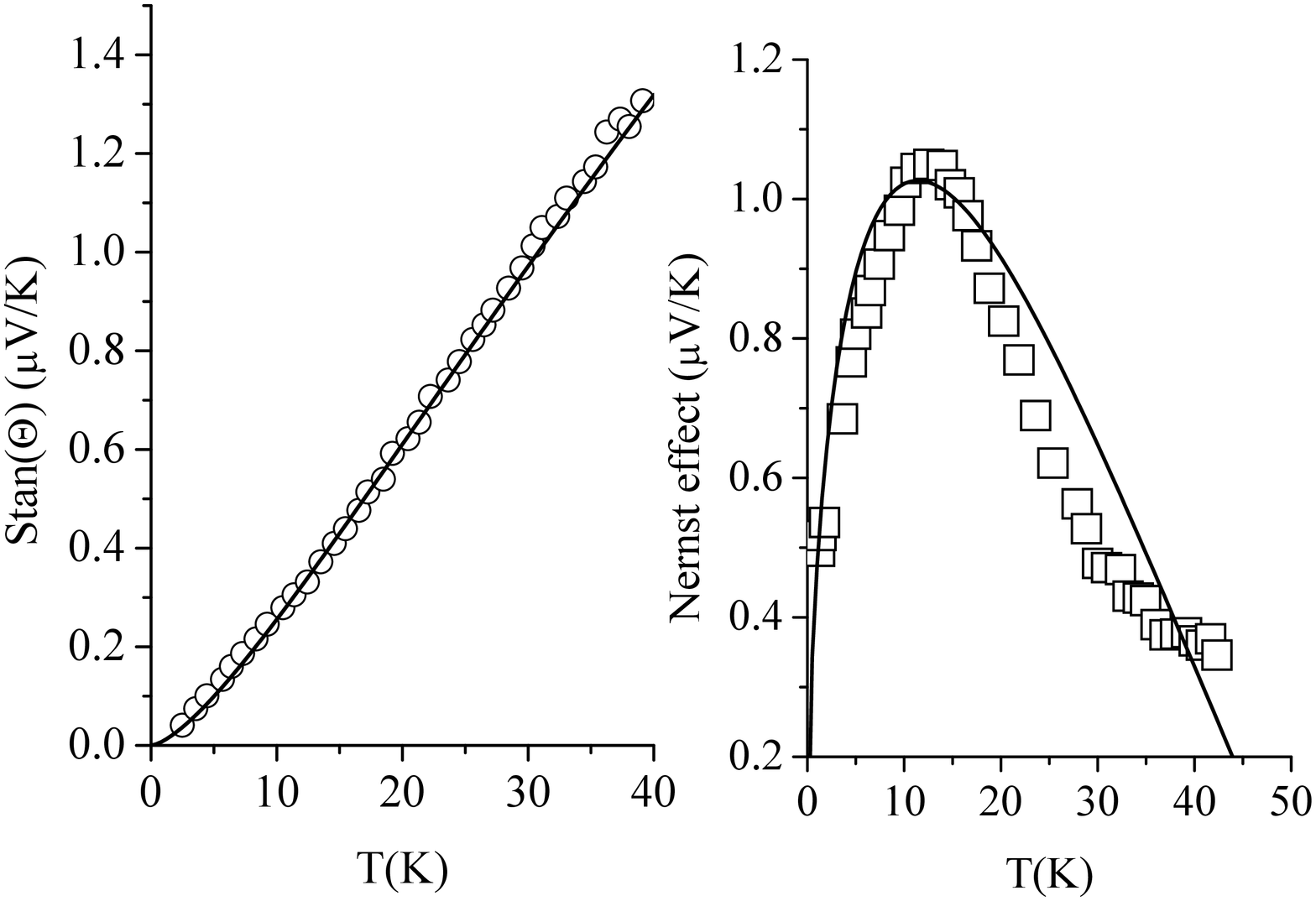}
\vskip -0.5mm \caption{$S\tan\Theta$ (circles \cite{cap2} )  and the
Nernst effect $e_y$  (squares \cite{cap})  of underdoped La$_{1.94}$
Sr$_{0.06}$CuO$_4$ at $B=12$ Tesla compared with the bipolaron
theory, Eqs.(18,19) (solid lines).}
\end{center}
\end{figure}

In spite of all simplifications, the model describes  remarkably
well both $S\tan \Theta$ and $e_y$  measured in La$_{1.94}$
Sr$_{0.06}$CuO$_4$ with a $single$ fitting parameter, $T_1=50$K
using the experimental $\rho(T)$. The constant  $e_0=2.95$ $\mu$V/K
scales the magnitudes of $S\tan \Theta$ and $e_y$.  The magnetic
field $B=12$ Tesla destroys the superconducting state of the
low-doped La$_{1.94}$ Sr$_{0.06}$CuO$_4$ down to $2$K, Fig.4, so any
residual superconducting order above $2$K is clearly ruled out,
while the Nernst signal, Fig.5, is remarkably large. The coexistence
of the large Nernst signal and a nonmetallic resistivity is in sharp
disagreement with the vortex scenario, but in agreement with our
model. Taking into account the field dependence of the conductivity
of localised carriers, the phonon-drug effect, and their
contribution to the transverse magnetotransport  can well describe
the magnetic field dependence of the Nernst signal \cite{alezav} and
improve the fit in Fig.5 at the expense of the increasing number of
fitting parameters.

\section{Normal state diamagnetism in cuprates}

A number of experiments (see, for example,
\cite{mac,jun,hof,nau,igu,ong} and references therein), including
torque magnetometries, showed enhanced diamagnetism above $T_c$,
which has been explained as the fluctuation diamagnetism in quasi-2D
superconducting cuprates (see, for example Ref. \cite{hof}). The
data taken at relatively low magnetic fields (typically below 5
Tesla) revealed a crossing point in the magnetization $M(T,B)$ of
most anisotropic cuprates (e.g. $Bi-2212$), or in $M(T,B)/B^{1/2}$
of less anisotropic $YBCO$ \cite{jun}. The dependence of
magnetization (or $M/B^{1/2}$) on the magnetic field has been shown
to vanish at some characteristic temperature below $T_c$. However
the data taken in high magnetic fields (up to 30 Tesla) have shown
that the crossing point, anticipated for low-dimensional
superconductors and associated with superconducting fluctuations,
does not explicitly exist in magnetic fields above 5 Tesla
\cite{nau}.

Most surprisingly the torque magnetometery  \cite{mac,nau} uncovered
a diamagnetic signal somewhat above $T_c$ which increases in
magnitude with applied magnetic field. It has been  linked with the
Nernst signal and mobile vortexes   in the  normal state of cuprates
\cite{ong}. However, apart from the inconsistences mentioned above,
the vortex scenario of the normal-state diamagnetism is internally
inconsistent.  Accepting the vortex scenario and fitting  the
magnetization data in $Bi-2212$  with the conventional  logarithmic
field dependence \cite{ong}, one obtains surprisingly high upper
critical fields $H_{c2} > 120$ Tesla and a very large
Ginzburg-Landau parameter, $\kappa=\lambda/\xi >450$  even at
temperatures close to $T_c$. The in-plane low-temperature magnetic
field penetration depth is $\lambda=200$ nm in optimally doped
$Bi-2212$ (see, for example \cite{tal}). Hence the zero temperature
coherence length $\xi$ turns out to be about  the lattice constant,
$\xi=0.45$nm, or even smaller. Such a small coherence length rules
out the "preformed Cooper pairs"  \cite{kiv}, since the pairs are
virtually not overlapped at any size of the Fermi surface in
$Bi-2212$ . Moreover the magnetic field dependence of $M(T,B)$ at
and above $T_c$ is entirely inconsistent  with what one expects from
a vortex liquid.  While $-M(B)$  decreases logarithmically at
temperatures well below $T_c$, the  experimental curves
\cite{mac,nau,ong} clearly show that   $-M(B)$  increases with the
field at and  above $T_c$ , just opposite to what one  could expect
in the vortex liquid.  This significant departure from the London
liquid behavior clearly indicates that the vortex liquid does not
appear above the resistive phase transition \cite{mac}.

Some time ago we  explained the anomalous diamagnetism in cuprates
as the Landau normal-state diamagnetism of preformed bosons
\cite{den}. The same model  predicted  the unusual upper critical
field \cite{aleH} observed in many superconducting cuprates
\cite{buc,mac0,boz,fra,gan,alezavnev,ZAV}. Here we extend the model
to high magnetic fields taking into account the magnetic
pair-breaking of singlet bipolarons and the  anisotropy of the
energy spectrum.

When the strong magnetic field is applied perpendicular to the
copper-oxygen plains the quasi-2D bipolaron energy spectrum is
quantized as
\begin{equation}
E_{\alpha}= \omega(n+1/2) +2t_c [1-\cos(k_zd)],
\end{equation}
where $\omega=2eB/m_b$, $n=0,1,2,...$, and $t_c$, $k_z$, $d$ are the
hopping integral, the momentum and the lattice period perpendicular
to the planes. Quantum numbers $\alpha$ also include  the momentum
along one of the in-plane directions. Expanding the Bose-Einstein
distribution function in powers of $exp[(\mu-E_\alpha)/T]$ with the
negative $\mu$ one can readily obtain (after summation over $n$) the
boson density
\begin{equation}
n_b={eB\over{\pi d}} \sum_{k=1}^{\infty} I_0(2t_c k/T) {\exp[ (\tilde{\mu} -2t_c)k/T]\over{1-\exp(-\omega k/T)}},
\end{equation}
and the magnetization
\begin{eqnarray}
M(T,B)&=&-n_b \mu_b+{eT\over{\pi d}} \sum_{k=1}^{\infty} I_0(2t_c
k/T) {\exp[ (\tilde{\mu} -2t_c)k/T]\over{1-\exp(-\omega k/T)}}
\nonumber \\
&\times& \left({1\over{k}}-{\omega \exp(-\omega
k/T)\over{T[1-\exp(-\omega k/T)]}}\right),
\end{eqnarray}
where $\mu_b=e/m_b$, $\tilde{\mu}=\mu-\omega/2$ and $I_0(x)$ is the
modified Bessel function. At low temperatures $T \rightarrow 0$
Schafroth's result \cite{sha} is recovered, $M(0,B)= -n_b \mu_b$.
The magnetization of charged bosons is field-independent at low
temperatures. At high temperatures, $T \gg T_c$ the chemical
potential has a large  magnitude , so we can keep only terms with
$k=1$ in Eqs.(21,22) to obtain
\begin{equation}
M(T,B)=-n_b \mu_b+{Tn_b\over{B}}  \left(1-{\omega \exp(-\omega /T)\over{T[1-\exp(-\omega/T)]}}\right).
\end{equation}
 The experimental conditions are such that $T \gg \omega$ when $T$ is of the order of $T_c$ or higher, so that
 \begin{equation}
M(T,B)=-n_b \mu_b {\omega\over{6T}},
\end{equation}
 which is the Landau  orbital diamagnetism  of nondegenerate carriers.
The  bipolaron in-plane mass in cuprates is about $m_b\approx 10
m_e$ \cite{alebook}. Using this mass yields $M(0,B) \approx 2000$
A/m with the bipolaron density $n_b=10 ^{21}$ cm$^{-3}$. Then the
magnitude and the field/temperature dependence of $M(T,B)$ near and
above $T_c$ are about the same as experimentally  observed in Refs
\cite{nau,ong}. The pseudogap temperature $T^*$ depends on the
magnetic field predominantly because of the magnetic-field splitting
of the single-polaron band in Fig.1. As a result the bipolaron
density depends on the field (as well as on  temperature) near $T_c$
as
 \begin{equation}
 n_b(T,B)=n_b(T_c,0)\left[1+(T_c-T)/\tilde{T}_0 -(B/B_0)^\beta \right],
 \end{equation}
 where $\tilde{T}_0$ and $B_0$ are constants  depending on $T^*$, $\beta=2$
if the polaron spectrum is spin-degenerate, and $\beta=1$ if the
spin degeneracy is removed by the crystal field already in the
absence of the external field.

 Theoretical temperature and field dependencies of $M(T,B)$,  Eq.(22)
agree qualitatively  with the experimental curves in $Bi-2212$
\cite{nau,ong}, if the depletion of the bipolaron density, Eq.(25)
is taken into account. The  depletion of $n_b$ accounts for the
absence of the crossing point in $M(T,B)$ at high magnetic fields.
Nevertheless a quantitative fit to experimental $M(T,B)$ curves
using $\tilde{T}_0$ and $B_0$ as the fitting parameters is
premature. The experimental diamagnetic magnetization has been
extracted from the total magnetization assuming  that the normal
state paramagnetic contribution remains temperature-independent  at
all temperatures \cite{nau,ong}. This assumption is inconsistent
with a great number of NMR and the Knight shift measurements, and
even with the preformed Cooper-pair model itself. The Pauli
spin-susceptibility has been found temperature-dependent in these
experiments revealing a normal-state pseudogap, contrary to the
assumption. Hence the experimental diamagnetic $M(T,B)$
\cite{nau,ong} has to be corrected by taking into account the
temperature dependence of the spin paramagnetism at relatively low
temperatures.

\section{Phase  coexistence and  resistivity near the ferromagnetic  transition in manganites}

   Ferromagnetic oxides, in particular  manganese perovskites,  show  a huge magnetoresistance near
the ferromagnetic transition.  The resistivity change is so large
that it could not compare with any other forms of magnetoresistance.
The effect observed in these materials was therefore named
'colossal' magnetoresistance (CMR) to distinguish it from the giant
magnetoresistance observed in magnetic multilayers. The discovery
raised expectations of a new generation of magnetic devices, and
launched a frenetic scientific race to understand the cause of the
effect.  Significant progress has been made in understanding their
properties, but new questions have arisen.  The ferromagnetic
metal-insulator transition in manganites has long been thought as
the consequence of the so-called double exchange mechanism (DEX),
which results in a varying bandwidth of electrons in the Mn$^{3+}$
d-shell as a function of  temperature \cite{dex}. More recently it
has been noticed \cite{mil} that the effective spin-exchange
interaction of the double-exchange model cannot  account for CMR
alone. In fact there is  strong experimental evidence for
exceptionally strong e-ph interactions in doped manganites from the
optical data (see section 2), the giant isotope effect \cite{zhao1},
the    Arrhenius behaviour of the drift and Hall mobilities
\cite{emi0} in the paramagnetic phase above the Curie temperature,
$T_m$, etc. Therefore Ref. \cite{mil} and  some subsequent
theoretical studies combined DEX with the Jahn-Teller e-ph
interaction in d-orbitals arriving at the conclusion that  the
low-temperature ferromagnetic phase is a spin-polarised metal, while
the paramagnetic phase is a polaronic insulator.

   However, some low-temperature optical \cite{opt},
electron-energy-loss (EELS) \cite{eels}, photoemission\cite{arpes}
and thermoelectric \cite{thermo} measurements showed that the
ferromagnetic phase of manganites is not a conventional  metal. In
particular, broad incoherent spectral features and a pseudo-gap in
the excitation spectrum were observed. EELS confirmed that
manganites were charge-transfer  doped insulators having p-holes as
 current carriers rather than d Mn$^{3+}$ electrons.
Photoemission and x-ray absorption spectroscopies of
La$_{1-x}$Sr$_x$MnO$_3$ also showed that the itinerant holes doped
into LaMnO$_3$ are  of oxygen p-character. CMR has been observed in
the ferromagnetic pyrochlore manganite Tl$_2$Mn$_2$O$_7$ \cite{ram},
which has neither the  mixed valence for DEX magnetic interaction
nor the Jahn-Teller cations such as Mn$^{3+}$.

   \begin{figure}
\begin{center}
\includegraphics[angle=-0,width=0.50\textwidth]{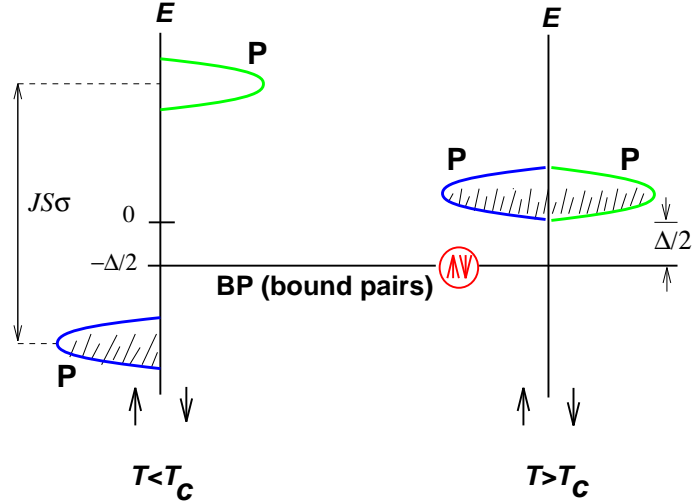}
\vskip -0.5mm \caption{Bipolaron model of CMR:  pairs (BP) are
localised on  impurity levels in the paramagnetic phase, where the
only current carriers are single thermally excited polarons (here
$T_c$=$T_m$ is the Curie temperature). If the exchange interaction
$JS\sigma$ between p-hole polarons and ordered manganese spins
exceeds the pair binding energy $\Delta$ , the pairs break at
$T<T_c$ because  the spin-up polaron sub-band sinks abruptly below
the bipolaron level. The ferromagnetic state is a polaronic
conductor.}
\end{center}
\end{figure}

These and  other observations \cite{zhao2}, in particular the fact
that  some samples of ferromagnetic manganites manifest an
insulating-like optical conductivity  at all temperatures
\cite{emi}, clearly rule out DEX  as the mechanism of CMR. They led
us to a novel theory of ferromagnetic/paramagnetic phase transition
and CMR based on the so-called  current-carrier density collapse
(CCDC) \cite{alebra2}. In CCDC p-holes are bound into heavy
bipolarons above $T_m$ due to the Fr\"ohlich e-ph interaction,
Eq.(2). The resistivity peak and CMR are the result of the magnetic
pair-breaking below $T_m$, Fig.6, caused by the $p-d$ spin-exchange
interaction, $J_{pd}$, which described as
\begin{equation}
H_{pd}= -(2N)^{-1} \sum_{\bf n,m} J_{pd}\hat{S}^z_{\bf
m}(c^{\dagger}_{\bf n \uparrow}c_{\bf n \uparrow}-c^{\dagger}_{\bf n
\downarrow}c_{\bf n \downarrow}).
\end{equation}
Here $\hat{S}^z_{\bf m}$ is the z-component of $Mn^{3+}$ spin on
site ${\bf m}$, and $N$ is the total number of sites.

Different from cuprates hole bipolarons are much heavier in
manganites because the e-ph Fr\"ohlich interaction is stronger and
the band structure is less anisotropic. They are readily localised
by disorder, so only thermally excited single extended polarons
conduct in the paramagnetic phase. With temperature lowering single
polarons polarize manganese spins at $T_m$ via $J_{pd}$, and the
spin polarization of manganese ions breaks the bipolaronic singlets
creating a spin-polarized $polaronic$ conductor. CCDC  explained the
resistivity peak and CMR in the experimental range of  external
magnetic fields \cite{alebra2,tai}. More recently, the theory has
been further confirmed  experimentally.  In particular, the oxygen
isotope effect has been observed in the low-temperature resistivity
of La$_{0.75}$Ca$_{0.25}$MnO$_3$ and Nd$_{0.7}$Sr$_{0.3}$MnO$_3$ and
explained by CCDC with polaronic carriers in the ferromagnetic phase
\cite{alezhao}. The current-carrier density collapse has been
directly observed using the Hall data in
La$_{0.67}$Ca$_{0.33}$MnO$_3$ and La$_{0.67}$Sr$_{0.33}$MnO$_3$
\cite{hall}. And the first order phase transition at $T_m$,
predicted by  the theory \cite{alebra2}, has been firmly established
in the specific heat measurements \cite{phil}. On the other hand,
the resistivity and the magnetization of some samples of
La$_{0.7}$Ca$_{0.3}$Mn$_{1-x}$Ti$_{x}$O$_{3}$ showed a more gradual
(second-order like) transition \cite{china}. Also the coexistence of
ferromagnetic and paramagnetic phases near the Curie temperature
observed in tunneling \cite{tun} and other experiments has not yet
been  addressed in the framework of CCDC. Here we argue that the
diagonal disorder, which is  inevitable with doping,  explains both
the phase coexistence and the resistivity/magnetization shape near
the transition.
\begin{figure}
\begin{center}
\includegraphics[angle=-90,width=0.55\textwidth]{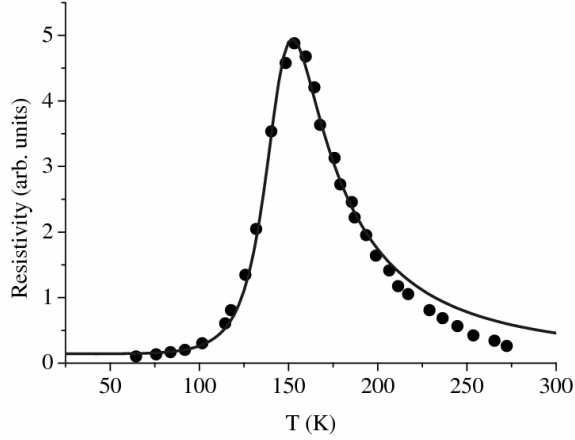}
\vskip -0.5mm \caption{CCDC model (Eq.(29), solid line) fits the
experimental
 resistivity near  ferromagnetic transition
in La$_{0.7}$Ca$_{0.3}$Mn$_{1.95}$Ti$_{0.05}$O$_{3}$ \cite{china}
(dots), when the phase coexistence caused by disorder is taken into
account.}
\end{center}
\end{figure}
The mean-field equations \cite{alebra2} describing the single
polaron density $n$, p-hole polaron $m$ and manganese $\sigma$
magnetizations, and the chemical potential $\mu=T \ln y$ can be
easily generalized by taking into account the random distribution of
the bipolaron binding energy $\delta= \Delta/(2J_{pd})$ across the
sample as
\begin{eqnarray}
&&n_i= 6y \cosh (\sigma_i/t), \cr
&& m_i= n_i \tanh (\sigma_i/t), \cr
&& \sigma_i= B_2 (m_i/2t), \cr
&& y^2= {{x-n_i}\over{18}}\exp(-2\delta_i/t),
\end{eqnarray}
where $t=T/J_{pd}$ is the reduced temperature, $B_S$ is the
Brillouin function, $x$ is the number of delocalised holes at zero
temperature in p-orbital states, which are   $3$-fold degenerate.
The subscript $i$ means different parts of the sample with different
$\delta_i$ because of disorder. While averaging these transparent
equations over a random distribution of $\delta_i$ is rather
cumbersome, one can apply a simplified approach using the fact that
the  phase transition in a homogeneous system is of the first order
in a wide range of $\delta$ \cite{alebra2}. Taking
$\sigma_i\approx=\Theta(T_{mi}-T)$ and $n_i\approx=x
\Theta(T_{mi}-T)+ \sqrt{2x} \exp(-\Delta/(2T))\Theta(T-T_{mi})$ and
averaging both quantities with the Gaussian distribution of random
$T_{mi}$s around the experimental $T_m$ we obtain the averaged
manganese magnetization
\begin{equation}
\sigma(T)={1\over{2}} erfc \left({{T-T_m}\over{\Gamma}}\right)
\end{equation}
and the resistivity, $\rho \propto 1/n$, near the transition
\begin{eqnarray}
1/\rho(T)&\propto& erfc\left({{T-T_m}\over{\Gamma}}\right)\cr
&+&(2/x)^{1/2} e^{-\Delta/2T}
erfc\left({{T_m-T}\over{\Gamma}}\right).
\end{eqnarray}
Here $\Delta$ is the average bipolaron binding energy, $\Theta(y)=1$
for $y>0$ and zero for $y<0$ , and $erfc(y)= (2/\pi^{1/2}) \int
_y^{\infty} dy \exp(-y^2)$.
 CCDC with disorder, Eq.(29) fits nicely the experimental
resistivity \cite{china} near the transition with physically
reasonable parameters $\Gamma=28$K,  $\Delta=1600$K, $T_m=102$K, and
$x=0.1$, Fig.7. A random distribution of transition temperatures
with the width $\Gamma$ across the sample caused by the randomness
of the bipolaron binding energy is responsible for the phase
coexistence near the transition \cite{tun}.

In summary, the strong-coupling bipolaron extension of the BCS
theory accounts for the kinetic properties of superconducting
cuprates including the temperature-dependent spin susceptibility,
the nonlinear in-plane resistivity, the maximum in the Hall effect,
the normal-state Nernst signal and the diamagnetism near and above
$T_c$. CMR and ferromagnetism of ferromagnetic oxides can be well
explained by the current-carrier density collapse in the framework
of the same theory including the exchange magnetic interaction of
p-holes with the manganese spins and disorder effects.

  I thank A.M. Bratkovsky, J.P. Hague, V.V. Kabanov, P.E. Kornilovitch,
J.H. Samson, P.E. Spencer, and V.N. Zavaritsky for collaboration and
valuable discussions. The work was supported by  EPSRC (UK) (grant
EP/C518365/1).}

\end{document}